\documentclass[preprint,12pt]{elsarticle}

\usepackage[tbtags]{amsmath}
\usepackage{bm}
\usepackage{amsfonts}
\usepackage{booktabs}
\usepackage{adjustbox}
\usepackage{graphics}
\usepackage{graphicx}
\usepackage{psfrag}
\usepackage{eurosym}
\usepackage{pstricks}
\usepackage{array}
\usepackage{shortvrb}
\usepackage{epsf}
\usepackage{graphicx}
\usepackage{rotating}
\usepackage{float}
\usepackage{color}
\usepackage{tabularx}
\usepackage{multirow, makecell}
\usepackage{float}
\usepackage{colortbl}
\usepackage{ifpdf}
\usepackage{multicol}
\usepackage{todonotes}
\usepackage[linesnumbered,ruled,vlined]{algorithm2e}

\usepackage{hyperref}
\hypersetup{colorlinks=true, linkcolor=blue, anchorcolor=red, citecolor=blue, filecolor=red, urlcolor=red, pdfauthor=author}
            
\usepackage{subfigure}

\usepackage{geometry}
\usepackage{alphalph}
\usepackage{etoolbox}
\usepackage{pdfpages}

\usepackage[T1]{fontenc}

\usepackage{caption}
\usepackage{subcaption}

\SetKwInput{KwInput}{Input}                
\SetKwInput{KwOutput}{Output}              

\newcommand{\comment}[1]{}

\makeatletter
\def\munderbar#1{\underline{\sbox\tw@{$#1$}\dp\tw@\z@\box\tw@}}
\makeatother

\patchcmd{\subequations}{\alph{equation}}{\alphalph{\value{equation}}}{}{}
\def \Noin {{\hskip 3pt \rm /\kern -9pt \in\hskip 1pt}}

\def \R {{\rm I\kern -2.2pt R\hskip 1pt}}

\allowdisplaybreaks

\begin{document}
\begin{frontmatter}

\title{Leveraging Neural Networks to Optimize Heliostat Field Aiming Strategies in Concentrating Solar Power Tower Plants}

\author[inst1]{Antonio Alc\'antara\corref{cor1}}
\ead{antalcan@est-econ.uc3m.es}
\author[inst1]{Pablo Diaz-Cachinero}
\ead{padiazc@est-econ.uc3m.es}
\author[inst2]{Alberto S\'anchez-Gonz\'alez}
\ead{asgonzal@ing.uc3m.es}
\author[inst1]{Carlos Ruiz}
\ead{caruizm@est-econ.uc3m.es}
\cortext[cor1]{Corresponding author}

\affiliation[inst1]{organization={Universidad Carlos III of Madrid, Department of Statistics}
            }

\affiliation[inst2]{organization={Universidad Carlos III of Madrid, Department of Thermal and Fluids Engineering, Energy Systems Engineering (ISE) Research Group}}

\begin{abstract}
Concentrating Solar Power Tower (CSPT) plants rely on heliostat fields to focus sunlight onto a central receiver. Although simple aiming strategies, such as directing all heliostats to the receiver’s equator, can maximize energy collection, they often result in uneven flux distributions that lead to hotspots, thermal stresses, and reduced receiver lifetimes. This paper presents a novel, data-driven approach that integrates constraint learning, neural network-based surrogates, and mathematical optimization to overcome these challenges. The methodology learns complex heliostat-to-receiver flux interactions from simulation data, constructing a surrogate model that is embedded into a tractable optimization framework. By maximizing a tailored quality score that balances energy collection and flux uniformity, the approach yields smoothly distributed flux profiles and mitigates excessive thermal peaks. An iterative refinement process, guided by the trust region and progressive data sampling, ensures the surrogate model improves the obtained solution by exploring new spaces during the iterations. Results from a real CSPT case study demonstrate that the proposed approach surpasses conventional heuristic methods, offering flatter flux distributions and safer thermal conditions without a substantial loss in overall energy capture.
\end{abstract}

\begin{keyword}
Aiming Strategies \sep  Concentrating Solar Power Tower Plants \sep Constraint Learning \sep Neural Networks \sep Mixed-Integer Programming.
\end{keyword}

\end{frontmatter}

\section{Introduction}
\label{sec:intro}

In Concentrating Solar Power Tower (CSPT) plants, a field of heliostats reflects and concentrates direct solar radiation onto a central receiver positioned on top of a tower \cite{vant2012central}. Although aiming every heliostat at the receiver’s center (single-point aiming) maximizes solar energy collection by minimizing spillage, multi-point aiming strategies are needed to mitigate thermo-mechanical constraints such as stress, fatigue, and corrosion within the receiver \cite{Sanchez-Gonzalez2016a}.

Conventional approaches to multi-point aiming typically rely on predefined rules tailored to specific geometries. Examples include the ``image size priority'' strategy for flat receivers \cite{Watkins2020}, the ``modified deviation-based aiming'' method for sodium receivers \cite{Wang2021}, and the ``aiming factor'' approach for cylindrical \cite{Sanchez-Gonzalez2018} and circular receivers \cite{Erasmus2022}.

However, rule-based strategies often fail to produce optimal aiming configurations. While binary integer linear programming \cite{Ashley2017a} and deterministic algorithms \cite{Sanchez-Gonzalez2024} have been explored, their computational costs hinder real-time implementation. As a result, researchers have turned to metaheuristics. For instance, \cite{Besarati2014} applied a genetic algorithm to a flat plate receiver, \cite{Yu2014} employed a Tabu search for a cavity receiver, and \cite{Flesch2017a} utilized ant colony optimization for a cylindrical receiver. These methods achieve more uniform flux distributions and minimize spillage, but their performance remains constrained by the complexity of the underlying optimization tasks.

Recent advancements in Artificial Intelligence (AI) and Machine Learning (ML) offer promising opportunities to enhance CSPT operation. Various studies have successfully applied AI and ML to different components of the CSPT system, such as the heliostat field, receiver, storage, and power block \cite{Milidonis2021, turja2024multi}, demonstrating encouraging results. For example, \cite{Ouyang2023} employed Neural Networks (NNs) to predict a plant’s power production. In terms of real-time aiming, \cite{Zeng2022} proposed a reinforcement learning strategy, \cite{Wu2024} developed a deep NN-based approach, and \cite{Carballo2025} introduced a deep reinforcement learning framework for the CESA-1 heliostat field. While these techniques adapt to changing conditions, they remain challenging to formulate and solve efficiently.

ML-based surrogate modeling presents a promising alternative for optimizing aiming strategies. Constraint Learning (CL), which integrates data-driven predictive models such as NNs, decision trees, or ensembles into optimization tasks, provides a structured framework for this hybrid approach \cite{fajemisin2024optimization, maragno2023mixed}. CL has found diverse applications in energy systems, including reducing computational burdens in biodiesel production \cite{fahmi2012process}, improving power system security \cite{cremer2018data, halilbavsic2018data}, regulating voltages in power grids \cite{chen2020input}, and verifying safe operational conditions \cite{venzke2020verification}. More recent contributions have used decision trees to improve tractability in bilevel programming \cite{prat2022learning}, embedded distributional models into tractable stochastic optimization \cite{alcantara2023neural}, incorporated nonconvex surrogates for optimal bidding \cite{dolanyi2023capturing}, applied Trust Region (TR) strategies to ensure solution reliability \cite{song2023constraint}, and used quantile models and distributional NNs to handle uncertainty and risk aversion \cite{alcantara2024optimal, alcantara2024quantile}. Advances in the efficiency and scalability of these approaches have been made by integrating ReLU-activated NNs and iterative learning \cite{zhou2024accelerating}.

Although CL has often been applied when the underlying constraints are unknown or implicit, the same principles can guide a hybrid approach for CSPT aiming. Here, we have theoretical models of the underlying physical phenomena, but these are too intricate to embed directly. Instead, we can leverage them to generate training data and then approximate the underlying physics using regression-based surrogates. This yields learned constraints that retain critical physical insights while making the problem tractable within an optimization framework. Despite the potential advantages of this approach, CL has not yet been explored in the CSPT aiming context, where complexity and nonlinearity are prevalent and could benefit significantly from data-driven optimization techniques.

In light of these considerations, this work makes several key contributions to the state-of-the-art in heliostat aiming strategies for CSPT plants:

\begin{itemize}
    \item We present the first integration of CL into CSPT aiming optimization, embedding complex, data-driven representations of heliostat-to-receiver flux interactions into a mathematical framework.
    \item By maximizing a tailored aiming strategy quality score ($QS$), the proposed method yields continuous and flexible aiming strategies that produce flatter, more uniform flux distributions on the receiver compared to standard $k$-factor sweep heuristics \cite{Sanchez-Gonzalez2018}.
    \item The proposed methodology affords a systematic trade-off between maximizing solar energy collection and mitigating thermal peaks, reducing the risk of hotspots that threaten receiver longevity.
    \item An iterative, trust-region-enriched solution procedure, combined with progressive sampling, ensures that the NN surrogate remains accurate in the most relevant regions of the design space, thereby improving solutions with each iteration.
    \item Although demonstrated on a particular CSPT configuration, the generalizability and adaptability of this approach make it suitable for a wide range of solar tower plant layouts and operating conditions.
\end{itemize}

The remainder of this manuscript is organized as follows. Section \ref{sec:background} outlines the relevant theoretical background and problem formulation. Section \ref{sec:methodology} describes the methodology, encompassing data generation, model training, and the associated optimization framework. Section \ref{sec:cases_estudies} presents and discusses the numerical results for our case study. Finally, Section \ref{sec:conclu} summarizes the key findings and proposes avenues for future research.

\section{Background}
\label{sec:background}

This section provides a basis for understanding the heliostat field aiming problem and the strategies employed to achieve optimal flux distributions on the receiver. We begin by reviewing a classical aiming approach, introducing key concepts such as the alternating top-bottom aiming pattern, $k$-factors, and vertical shifts that help tailor flux profiles to meet the receiver's thermomechanical constraints. We then discuss the underlying optical model used to generate flux maps, detailing the methods and parameters that influence how solar energy reflects and concentrates onto the receiver’s surface. Building on these foundations, we introduce the CL paradigm and describe how neural network-based surrogates can be embedded within an optimization framework. This provides a pathway to handle complexity and nonlinearity in real-world conditions, setting the stage for the developed methodology.

\subsection{Classic Aiming Strategy}\label{sec:classical_aiming}

\begin{figure}[!ht]
    \centering
    \includegraphics[width=0.4\linewidth]{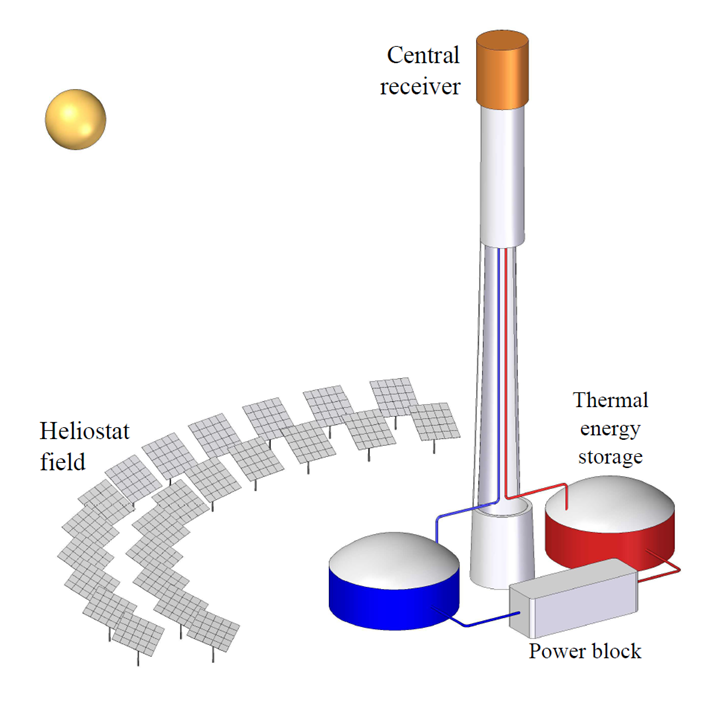}
    \caption{Schematic representation of a CSPT plant.}
    \label{fig:esquema_cspt}
\end{figure}

Figure \ref{fig:esquema_cspt} provides a simplified schematic representation of a CSPT plant, showcasing its principal components and overall configuration. In such a system, an array of heliostats (large, sun-tracking mirrors) is arranged around a central tower. Each heliostat concentrates sunlight onto a receiver mounted at the top of the tower. The receiver absorbs this concentrated solar energy, transferring it to a working fluid that is subsequently stored in thermal energy storage tanks. This stored energy can then be directed to a power block, where it is converted into electricity on demand, providing a dispatchable renewable energy source.

As presented in the introduction, a multi-point aiming strategy is essential to meet the receiver’s requirements. \cite{Vant-Hull2002} introduced an aiming strategy in which each row of heliostats is aimed alternately at the top and bottom of cylindrical receivers to produce flux distributions that are symmetrical with respect to the receiver equator. Specifically, each heliostat is aimed so that its beam is tangent to the top or bottom edge of the receiver.

In the present work, heliostats in the odd-numbered rows are aimed at the upper half, and heliostats in the even-numbered rows are aimed at the lower half. Accordingly, the aim point vertical shift with respect to the receiver equator ($y_{shift}$) follows \eqref{eqYshift} where $RH$ and $R_k$ respectively stand for the receiver height and the heliostat beam radius.

\begin{equation}
{y_{shift}} = \left\{ {\begin{array}{*{20}{c}}
{ + \left( {{RH}/{2} - {R_k}} \right),}&{{R_k} < {RH \mathord{\left/
 {\vphantom {H 2}} \right.
 \kern-\nulldelimiterspace} 2}} \quad \& \quad odd \; row\\
{ - \left( {{RH}/{2} - {R_k}} \right),}&{{R_k} < {RH \mathord{\left/
 {\vphantom {H 2}} \right.
 \kern-\nulldelimiterspace} 2}} \quad \& \quad even \; row\\
0, &{{R_k} \ge {RH \mathord{\left/
 {\vphantom {H 2}} \right.
 \kern-\nulldelimiterspace} 2}}
\end{array}} \right.
\label{eqYshift}
\end{equation}

To estimate the beam radius, the concept of $k$ aiming factor was developed \cite{Sanchez-Gonzalez2018}. Such a $k$-factor typically ranges between 0 (aiming at the edge) and 3 (aiming at the equator). The heliostat beam radius is calculated as a function of the $k$-factor using \eqref{eq.R} where $S$ stands for the slant range (heliostat-to-target distance), $\epsilon_r$ denotes the incident angle on the receiver and $\sigma_e$ stands for the effective error. The effective error results from the convolution of sunshape ($\sigma_{sun}$), heliostat slope error ($\sigma_{slp}$), and tracking error ($\sigma_{trk}$) as reported in \eqref{eq.sigma}, where $\omega$ represents the heliostat incidence angle. For the sunshape standard deviation, $\sigma_{slp}$ = 2.09 mrad is a generally adopted value~\cite{biggs1979helios,Schmitz2006111}.

\begin{equation}
R_k = \frac{ k \cdot \sigma_{e} \cdot S} { \cos \varepsilon_r }
\label{eq.R}
\end{equation}

\begin{equation}
{\sigma _e} = \sqrt {\sigma _{sun}^2 + 2\left( {1 + \cos {\omega}} \right)\sigma _{slp}^2 + \sigma _{trk}^2} 
\label{eq.sigma}
\end{equation}

Uniform flux distributions are key to reduce thermal stress and ensure the receiver’s lifetime \cite{Logie2018}. Therefore, the aiming strategy must result in flat flux profiles on the receiver panels, while minimizing energy losses.

To flatten the flux profiles, a classic aiming strategy was based on sweeping the $k$-factors. From $k$=3, a downward sweep is performed so that the largest central peak (equatorial aiming) is gradually reduced. As the $k$-factor decreases, the profile flattens and, ultimately, two peaks (shoulders) emerge. For each heliostat field sector, the $k$-factor right before the appearance of the two peaks was selected. For more details on the $k$-factor concept and the sweep approach, see reference \cite{Sanchez-Gonzalez2018}.

\subsection{Flux mapping model}

The optical model generates the flux maps on the receiver, considering any heliostat field aiming. The flux mapping model used in this study is based on the Convolution-Projection (CP) method previously presented in \cite{Sanchez-Gonzalez2015}. The receiver mesh is projected into the image plane, where an analytic function is evaluated. Such analytic function relies on the convolution of sunshape, concentration and optical errors, as the expression in \eqref{eq.sigma}, based on UNIZAR~\cite{Collado1986}. 

The flux maps are expressed in terms of the concentration ratio of flux density ($C^M_{p,v,h}$) at any node in the receiver, identified by the vertical ($v$) and horizontal ($h$) position in the panel ($p$). Such a concentration ratio is defined as the ratio of flux density to the direct normal irradiance ($C^M_{p,v,h} = F^M_{p,v,h} / DNI$). Thus, $C^M_{p,v,h}$ represents the number of suns impinging on any point in the receiver, regardless of the instantaneous $DNI$. 

The model computes all optical losses in the field, such as cosine, attenuation, shading, blocking, and spillage. A metric of interest in this sense is the spillage loss factor ($SPL$), which represents the fraction of solar energy reflected by the heliostat field that is not intercepted by the receiver. The spillage loss factor is calculated using \eqref{eq.SPL}, where $AM$ and $dA$ are the heliostat mirror area and the receiver area covered by a node, respectively. 

\begin{equation}
    SPL = 1 - \frac{\textstyle \sum_{p=1}^{P} \sum_{v=1}^{V} \sum_{h=1}^{H} C^M_{p,v,h} \cdot dA} {\textstyle \sum_{i=1}^{N^{heliostats}} \cos{\omega_i} \cdot AM}
    \label{eq.SPL}
\end{equation}


    \label{eq.CV}

\subsection{Constraint Learning and Model Embedding}\label{sec:constraint_lear}

CL is a powerful methodology that integrates ML and optimization to address decision-making problems where constraints and objectives are not explicitly defined, are difficult to express analytically, or are inherently nonlinear \cite{maragno2023mixed}. By leveraging historical or simulated data, CL approximates complex relationships between decision variables and outcomes through predictive models, such as NNs \cite{gallant1993neural}, decision trees \cite{breiman1984classification}, or ensemble methods \cite{breiman2001random} like random forests and gradient-boosting machines. These models can be embedded within optimization frameworks as Mixed-Integer Linear Programming (MILP) formulations, enabling the incorporation of learned constraints and objective functions  \cite{fajemisin2024optimization}. This approach facilitates solving problems where traditional optimization methods fail, allowing the inclusion of non-quantifiable constraints and dynamic system behavior. By bridging the gap between data-driven insights and optimization, CL stands as a powerfull framework for tackling problems with high-dimensional, nonlinear relationships.

Suppose we have access to a dataset $\textbf{D} = \{(x_i, \omega_i, y_i)\}_{i=1}^N$, where $x_i$ represents the decision variables,  $\omega_i$ denotes the contextual information, and $y_i$ is the outcome of interest for sample $i$. The goal of CL is to use the dataset $\textbf{D}$ to train predictive models $\hat{h}_{\textbf{D}}(x, \omega)$, which approximate the relationship between decisions, context, and outcomes. These models can then be embedded into optimization problems to constrain or optimize $y$ for new contexts $\omega$ and decisions $x$.

Mathematically, this can be represented as:
\begin{subequations}
\begin{align}
    \min_{x \in \mathbb{R}^{n_0}, y \in \mathbb{R}^k} & \quad f(x, \omega, y) \\
    \text{s.t.} & \quad g(x, \omega, y) \leq 0, \\
                & \quad y = \hat{h}^{\textbf{D}}(x, \omega), \\
                & \quad x \in X(\textbf{D}),
\end{align}
\end{subequations}

where $f(x, \omega, y)$ is the objective function, $g(x, \omega, y)$ represents explicit constraints, $\hat{h}_\textbf{D}(x, \omega)$ is the learned predictive model trained on $\textbf{D}$, and $X(\textbf{D})$ defines the TR. The TR allows us to limit decisions $x$ to be around the Convex Hull (CH) of samples in $\textbf{D}$, which were already used during the training of $\hat{h}_{\textbf{D}}(x, \omega)$. This helps to ensure the validity of the embedded models.

In the context of CSPT plants, CL offers a suitable approach to optimize heliostat field aiming strategies. The constraints in this application involve intricate interactions between heliostat configurations (decision variables $x$), receiver interception, and spillage losses (response variables $y$). While theoretical equations can model these relationships, solving them within an optimization problem remains challenging due to their nonlinear nature. Using historical or simulated operational data, predictive models can learn these interactions and embed them into optimization frameworks, enabling precise control over heliostat aim points. This data-driven methodology can enhance energy collection, reduce receiver thermal stresses and spillage losses, while improving the overall plant performance. By tailoring aiming strategies to the specific geometry and operational parameters of each plant, CL can surpass traditional heuristics, offering a more adaptable and detailed framework for optimizing CSPT plant operations.

\subsection{Mixed-integer Formulation of ReLU Networks} \label{sec:mip_relu}

NNs offer greater flexibility for modeling intricate nonlinear relationships, especially in high-dimensional data, by approximating the conditional mean of the response variable given predictors and minimizing residuals, which enables effective learning and generalization from observed data \cite{gallant1993neural}. Classical NNs have a well-established architecture consisting of an input layer that receives predictors, one or more hidden layers that transform data through fully connected layers with nonlinear activation functions (e.g., ReLU, Tanh and Sigmoid), possibly using regularization techniques like dropout to prevent overfitting, and an output layer that produces predictions, which may represent single or multiple values depending on the task.

Mathematically, an $ L $-layer classical NN \cite{Goodfellow-et-al-2016} is defined with several components. Firstly, an input $ \mathbf{X} \in \mathbb{R}^{n_x} $, where $ n_x $ is the number of features. Each layer $ l \in \{1, \dots, L\} $ has a weight matrix $ W^{l} \in \mathbb{R}^{n^{l} \times n^{l-1}} $ and a bias vector $ b^{l} \in \mathbb{R}^{n^{l}} $, where $ n^{l} $ is the number of neurons in layer $ l $ (with $ n^{0} = n_x $). The activation function for layer $ l $ is $ g^{l}(\cdot): \mathbb{R} \to \mathbb{R} $. The output of layer $ l $ is computed as:
\begin{subequations}
\begin{align}
z^{l} &= W^{l} a^{l-1} + b^{l}, \\
a^{l} &= g^{l}(z^{l}),
\end{align}
\end{subequations}
with $ a^{0} = \mathbf{X} $, and the network output is $ \mathbf{y} = z^{L} \in \mathbb{R}^{n_y} $, where $ n_y $ is the output dimensionality.

NNs with piecewise linear activation functions (e.g., ReLU and leaky ReLU) can be reformulated as an equivalent set of mixed-integer linear constraints. Therefore, while the NN approximates nonlinear relationships, its behavior can be partitioned into regions defined by linear or piecewise linear constraints, allowing it to be embedded into a MILP formulation that captures its learned relationships while maintaining tractability.

Once an NN is trained, its parameters $ W^{l} $ and $ b^{l} $ are fixed, making the network a static (piecewise) linear mapping if all activation functions $ g^{l} $ are (piecewise) linear. In this case, the NN can be reformulated by considering each activation function separately \cite{huchette2023deep}. The big-$ M $ method is commonly used to represent disjunctive constraints in mixed-integer programming \cite{bonami2015mathematical}, valued for its simplicity and compactness. However, its linear relaxations can be weak, potentially slowing solver performance. For complex optimization problems, stronger formulations providing tighter relaxations can enhance solver efficiency \cite{grimstad2019relu, anderson2020strong, tsay2021partition}.


Consider a single ReLU activation function in an NN layer $ l $ with $ n^{l} = 1 $, defined as:
\begin{align}
a^{l} = \max\{0, W^{l} a^{l-1} + b^{l}\}.
\end{align}

Early studies \cite{lomuscio2017approach, tjeng2017evaluating, fischetti2018deep} used the big-$ M $ method to derive a mixed-integer formulation for ReLU activations:
\begin{subequations}\label{eq:relu}
\begin{align}
a^{l} &\geq W^{l} a^{l-1} + b^{l}, \label{eq:relu1} \\
a^{l} &\leq W^{l} a^{l-1} + b^{l} - M^- (1 - \sigma), \label{eq:relu2} \\
0 \,\,\,\,\,&\leq a^{l} \leq M^+ \sigma, \label{eq:relu3} \\
\sigma &\in \{0, 1\}, \label{eq:relu4}
\end{align}
\end{subequations}
where $ \sigma $ is a binary variable, and $ M^+ $ and $ M^- $ are respectively positive and negative big-$ M $ constants satisfying:
\begin{align}
M^- \leq W^{l} a^{l-1} + b^{l} \leq M^+. \label{eq:bounds}
\end{align}

A common approach to select these big-M values in shorter, less deep NNs is using interval arithmetic. Given bounds $ a_i^{l-1} \in [\underline{a}_i^{l-1}, \overline{a}_i^{l-1}] $, valid $ M^- $ and $ M^+ $ can be computed as:
\begin{subequations}
\begin{align}
M^- &= \sum_i \underline{a}_i^{l-1} \max(0, W_i^{l}) + \overline{a}_i^{l-1} \min(0, W_i^{l}) + b^{l}, \label{eq:interval1} \\
M^+ &= \sum_i \overline{a}_i^{l-1} \max(0, W_i^{l}) + \underline{a}_i^{l-1} \min(0, W_i^{l}) + b^{l}. \label{eq:interval2}
\end{align}
\end{subequations}

By embedding the NN constraints directly into the optimization problem, it is possible to learn complex constraints from data and enforce them within the decision-making process. This synergy enhances the ability to model and solve complex real-world problems where learned constraints are essential. As mentioned before, in a CSPT-based optimization problem, we may need to set the aiming strategy in order to flatten the flux distribution on the receiver, while minimizing the spillage losses.

\section{Methodology}
\label{sec:methodology}

This section describes the methodology developed to optimize heliostat field aiming strategies, starting with a case study using the Dunhuang CSPT plant as a reference. We introduce a quantitative measure of performance, which balances energy capture and flux uniformity to ensure safe and efficient receiver operation. The framework leverages a neural network-based surrogate to approximate the complex heliostat-to-receiver flux interactions, enabling tractable optimization formulations while retaining physical fidelity. Data-driven sampling techniques generate the necessary training sets, and constraints derived from the plant’s geometric and operational parameters shape the feasible solution space. Finally, we incorporate an iterative refinement procedure to enhance model accuracy and solution quality, systematically improving the heliostat aiming strategy across successive optimization cycles.

\subsection{Case Study}
The strategy described in the current research is applicable to any solar power tower plant with a central receiver. To demonstrate the process and present the results, the Dunhuang CSPT plant is used as a case study. This 10~MW$_e$ facility is located in China at a latitude of 40.08~\textdegree N.

\begin{figure}[!ht]
    \centering
    \includegraphics[width=0.5\linewidth]{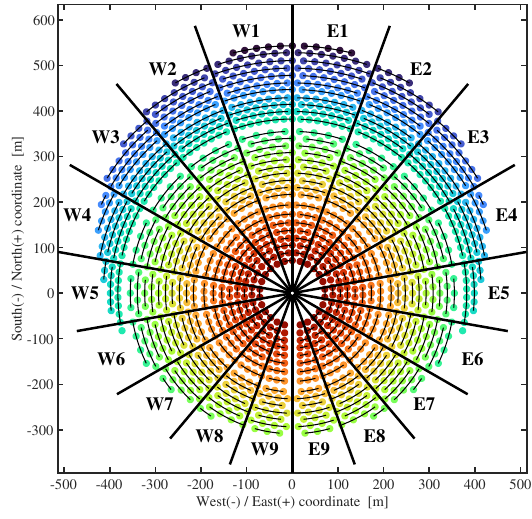}
    \caption{Heliostat field layout at Dunhuang 10~MW$_e$ plant. Rows are colored as a function of distance to the receiver for the sake of heliostat identification in the aiming maps.}
    \label{figHF}
\end{figure}

The field consists of 1525 square heliostats, arranged as shown in Figure~\ref{figHF}. The cylindrical receiver, located atop the tower, consists of 18 panels (9 in the East side, 9 in the West). Based on the 20\textdegree~aperture angle of the panels, the field is divided in 18 sectors, as labeled in Figure~\ref{figHF}. Each panel measures 9.2~x 1.3~m. Table~\ref{tabDH} displays further data of Dunhuang 10~MW$_e$ plant. For the calculation of \eqref{eq.sigma}, the mirror slope error is assumed to be 2.6~mrad, while $\sigma_{trk}$ is neglected.

\begin{table}[!ht]
  \centering
    \begin{tabular}{llr}
    \toprule
    \multirow{3}[2]{*}{\textbf{Field}} & Number of square heliostats & 1525 \\
          & Heliostat mirror area ($AM$), m$^2$ & 115.7 \\
          & Tower optical height, m & 121.4 \\
    \midrule
    \multirow{4}[2]{*}{\textbf{Receiver}} & Diameter ($RD$), m & 7.3 \\
          & Height ($RH$), m & 9.2 \\
          & Number of panels & 18 \\
          & Panel width, m & 1.29 \\
    \bottomrule
    \end{tabular}%
    \caption{Design data of Dunhuang 10~MW$_e$ plant.}
  \label{tabDH}%
\end{table}%

\subsection{Aiming Strategy Quality Score}\label{sec:aiming_q_score}

In this section, we develop the concept of aiming strategy quality score $QS$, inspired by the reward function developed by \cite{Carballo2025}. The calculation of this $QS$ serves the dual purpose of maximizing the energy collected on the receiver and ensuring uniform concentration distribution. Uniformity is critical to maintaining the structural integrity of the receiver and avoiding local overheating, which could lead to irreversible damage. This score is therefore a key component in the proposed framework.

\begin{figure}[!ht]
    \centering
    \includegraphics[width=\linewidth]{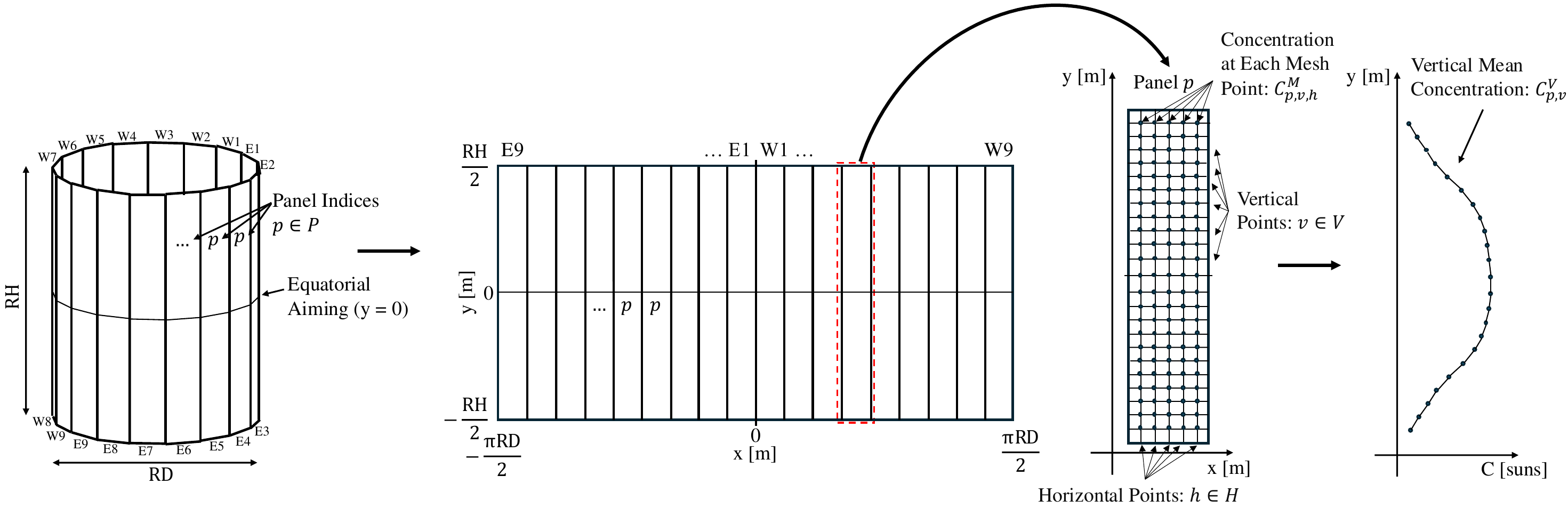}
    \caption{Receiver divided into $P$ panels, 2D representation of the panels in the receiver, and schematic representation of vertical mean concentration calculation.}
    \label{fig:AIMING_QUA_SCORE_2}
\end{figure}

Before computing the $QS$, note that the cylindrical receiver is approximated being divided into $P$ flat panels, as illustrated in Figure \ref{fig:AIMING_QUA_SCORE_2}. Moreover, each panel is discretized into vertical ($V$) and horizontal ($H$) mesh points, where the flux concentration $C^M_{p,v,h}$ is calculated at each mesh point with the CP method.

Algorithm \ref{algorithm_aiming_quality_score} illustrates the calculation of the $QS$. It begins by iterating over each panel $p \in P$ in the receiver (line \ref{linefor}). For each panel, the flux concentration values are computed using the CP method described in \cite{Sanchez-Gonzalez2015}, which considers the heliostat aiming factors $\textbf{X}$ and the contextual information $\boldsymbol{\Omega}$, such as solar position and geometrical data of the CSPT plant (line \ref{linecon}). Line \ref{lineconmedia} computes the vertical mean concentration $C^V_{p,v}$ by averaging the concentration values across all horizontal mesh points $h \in H$. Then, this mean concentration is normalized to obtain $S^V_{p,v}$, scaling the values between the minimum and maximum vertical concentrations in the panel (line \ref{lineconpu}). To evaluate uniformity, the distribution differences $dd_{p,v}$ are calculated for central vertical mesh points $v \in V^C$ as $1 - S^V_{p,v}$ (line \ref{linecondiff}), and the average difference $\overline{dd}_p$ for each panel is computed as the mean of these deviations (line \ref{linecondiffmean}). These steps ensure that uniformity in the central region of the panel is explicitly considered. 

Line \ref{lineenergy} allows the computation of the total energy $E_p$, which corresponds to the area under the vertical mean concentration curve, calculated using the \texttt{trapz} function from \texttt{Numpy} \cite{idris2015numpy}. The $QS$ for each panel, $QS^{\text{Panel}}_p$, is then obtained by subtracting a penalty term, proportional to $\overline{dd}_p$ weighted by $\lambda$, from the total energy $E_p$ (line \ref{lineqpanel}). This ensures that the score reflects both the energy collected and the uniformity of the flux distribution. After computing the scores for all panels, the overall $QS$ is determined by computing a weighted mean of the panel scores, where the weights correspond to the number of heliostats assigned to each panel $N^{\text{Heliostats}}_p$ (line \ref{linefinal}). This process ensures that the overall score trades-off between energy maximization and uniformity, controlling it with the penalty factor $\lambda$.

\begin{algorithm}[!ht]
\caption{Aiming strategy quality score calculation}
\label{algorithm_aiming_quality_score}

\KwInput{
    $C^M_{p,v,h}$: Flux concentration on the mesh for each panel, vertical position, and horizontal position \\
    $H$: Horizontal mesh points in the panel \\
    $N^{\text{Heliostats}}_p$: Number of heliostats per panel \\
    $P$: Panels in the receiver \\
    $V$: Vertical mesh points in the panel \\
    $V^C$: Vertical central mesh points in the panel\\
    $\textbf{X}$: Aiming factors $k$ for heliostats\\
    $\lambda$: Weighting factor for the distribution difference penalty \\
    $\boldsymbol{\Omega}$: Contextual and technical information for the CSPT plant\\
}

\KwOutput{
    $QS$: Overall aiming strategy quality score
}

\For{$p = 1$ \KwTo $P$\label{linefor}}{
    Compute $C^M_{p,v,h}$  using the CP method for the values of $\textbf{X}$ and considering  $\boldsymbol{\Omega}$ \label{linecon}
    
    $C^V_{p,v} = \langle C^M_{p,v,h} \rangle_{h \in H}$\label{lineconmedia}
    
    $S^V_{p,v} = \frac{C^V_{p,v} - \min_{v \in V} C^V_{p,v}}{\max_{v \in V} C^V_{p,v} - \min_{v \in V} C^V_{p,v}}$ \label{lineconpu}
    
    $dd_{p,v} = 1 - S^V_{p,v}$,  $v \in V^C$  \label{linecondiff}
    
    $\overline{dd}_p = \langle dd_{p,v} \rangle_{v \in V^C}$ \label{linecondiffmean}
    
    $E_p = \int_{v \in V} C^V_{p,v} \, \text{d}v$ using the \texttt{trapz} function \label{lineenergy}
    
    $QS^{\text{Panel}}_p = E_p - \lambda \cdot \overline{dd}_p$ \label{lineqpanel}
}

Save $QS = \frac{\sum_{p=1}^P QS^{\text{Panel}}_p \cdot N^{\text{Heliostats}}_p}{\sum_{p=1}^P N^{\text{Heliostats}}_p}$ \label{linefinal}
\end{algorithm}

\begin{figure}[!ht]
    \centering
    \includegraphics[width=.8\linewidth]{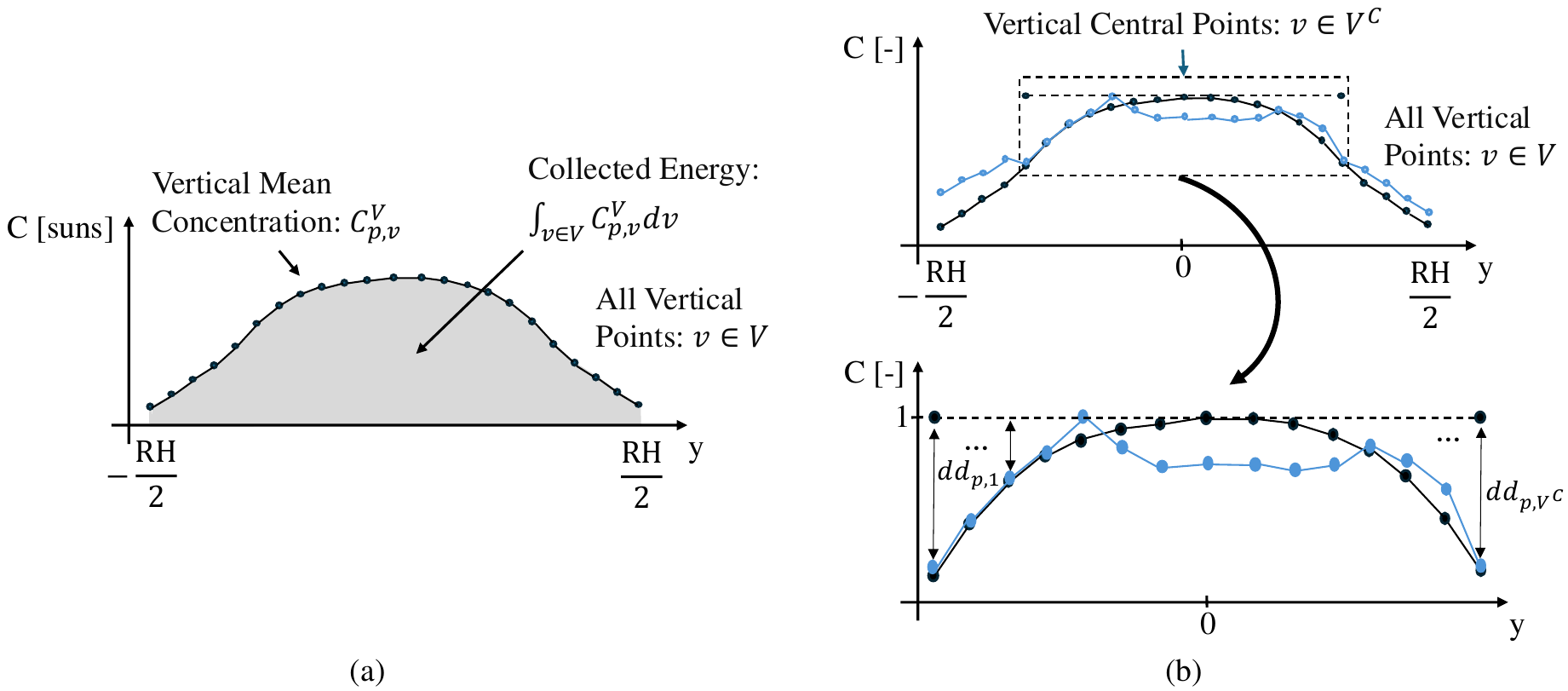}
    \caption{Schematic representation of aiming strategy quality score calculation: (a) Area calculation under the vertical mean concentration curve. (b) Penalty for deviation from uniform shape in the central part of the panel. }
    \label{fig:AIMING_QUA_SCORE_3}
\end{figure}

Figures \ref{fig:AIMING_QUA_SCORE_2} and \ref{fig:AIMING_QUA_SCORE_3} provides a schematic representation of these calculations. First, Figure \ref{fig:AIMING_QUA_SCORE_2} illustrates how the flux concentration is computed for each mesh point in a selected panel $p$, followed by the calculation of the vertical mean concentration $C^V_{p,v}$ by averaging the concentration values across all horizontal mesh points $h \in H$. Then, Figure \ref{fig:AIMING_QUA_SCORE_3} (a) illustrates the computation of the total collected energy $E_p$, which corresponds to the area under the vertical mean concentration curve $C^V_{p,v}$. Finally, Figure \ref{fig:AIMING_QUA_SCORE_3} (b) shows the calculation of the penalty for distribution difference from uniformity for two different vertical mean concentration curves (blue and black likes), emphasizing the importance of concentration consistency in the central region.

\subsection{Neural Network Architecture}

Our methodology leverages an NN as a surrogate model to optimize the aiming strategy of a CSPT plant, providing a learning-based alternative to heuristic techniques. The equations governing the aiming process involve nonlinear and nonconvex terms, posing significant challenges for optimization since current solvers often yield only local optima. While heuristic methods approximate solutions through exhaustive search within a limited scope, our approach integrates the surrogate model into a classical optimization framework, enabling a precise mathematical representation of the regression model. This integration guarantees the optimal solution of the surrogate problem and significantly improves the heliostat aiming strategy by overcoming the limitations of heuristic techniques. Additionally, our method delivers higher-quality solutions than those currently available in the literature, achieving better uniformity compared to the Sweep approach proposed in \cite{Sanchez-Gonzalez2018}, for instance.

\begin{figure}[!ht]
    \centering
    \includegraphics[width=0.8\linewidth]{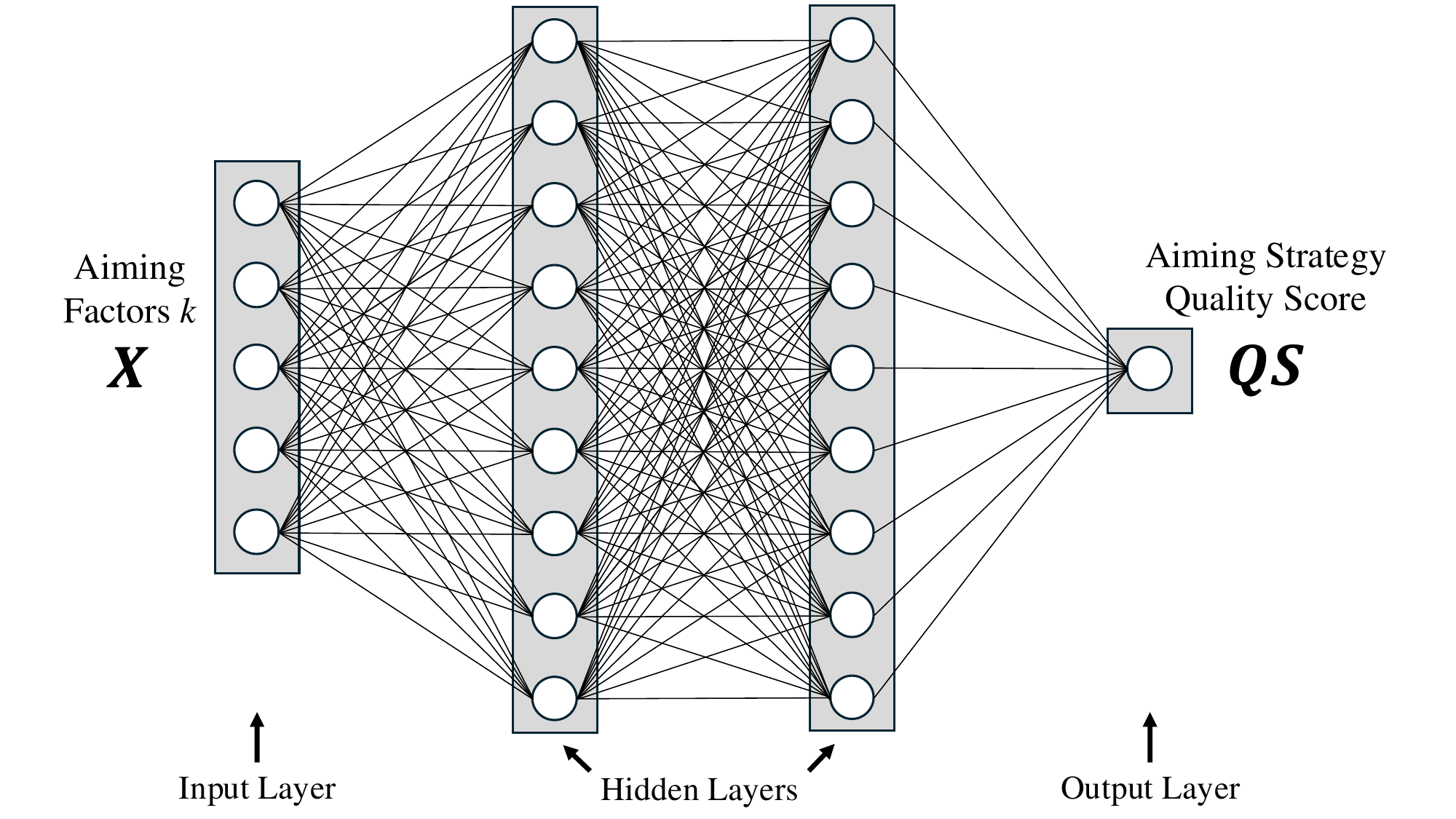}
    \caption{NN structure for optimizing heliostat field aiming strategies.}
    \label{fig:NN_archt}
\end{figure}

To achieve this goal, we train an NN to predict the $QS$ of a given aiming strategy, using the aiming factors $k$ as inputs. As illustrated in Figure \ref{fig:NN_archt}, the NN is designed as a single-output feedforward NN. The decision variables $\textbf{X}$, which represent the aiming factors, serve as inputs to the network, while the output layer generates the corresponding aiming strategy quality score $QS$. This score is a function of the aiming factors $\textbf{X}$ and the contextual information $\boldsymbol{\Omega}$, facilitating the reconstruction of the aiming strategy on the CSPT plant's receiver and providing a direct connection between the decision variables and the performance metric. Therefore, NN serves as a surrogate model for the quality score. To seamlessly integrate this NN into a classical optimization problem, all activation functions in the network are chosen to be piecewise linear, specifically employing ReLU for all nodes, as detailed in Section \ref{sec:mip_relu}. This choice enables the NN approximation of the aiming strategy to be expressed as a set of piecewise-linear constraints, ensuring efficient integration into the optimization framework.

\subsection{Data Generation}

Efficient data generation is essential for training the NN used to approximate the $QS$. Algorithm \ref{algorithm_pata_gen} describes the process of constructing the dataset $ D $, which consists of pairs $(\textbf{X}_i, QS_i)$, for $i=1,\dots,N$. Here, $\textbf{X}_i$ represents feasible random values of heliostat aiming factors, and $QS_i$ is their corresponding quality score value. This procedure generates the dataset required for the initial training iteration by producing random values of the aiming factors ($\textbf{X}_i$) and evaluating them to compute their quality scores ($QS_i$). In subsequent iterations, the process refines the data by generating random values around the optimal solution found by the NN in the previous iteration. This approach ensures a progressive focus on the most promising regions of the solution space, as will be detailed in Section \ref{sec_itera_procedure}. By leveraging random sampling and efficient evaluations, the dataset is prepared in a straightforward and computationally efficient manner.

\begin{algorithm}[!ht]
\KwInput{
$a, b$: Limits of Uniform distribution\\
$\mu, \sigma$: Mean and standard deviation of Normal distribution\\
$N$: Dataset size\\
$t$: Iteration index\\
$\textbf{X}^*$: Optimal solution from Problem~\eqref{problem_nn_final} if $t>1$
}
\KwOutput{$D = \{(\textbf{X}_1, QS_1), \ldots, (\textbf{X}_N, QS_N)\}$: Generated dataset}

Initialize $D \leftarrow \{\}$

\For{$i=1$ \KwTo $N$}{
    \If{$t = 1$}{
        Generate $\textbf{X}_i$ as feasible random values of $k$ using $N(\mu, \sigma)$ or $U(a, b)$ \label{linei1}
    }
    \Else{
        Generate $\textbf{X}_i$ as feasible random values around $\textbf{X}^*$ using $N(\textbf{X}^*, \sigma)$ or $U(\textbf{X}^*-a,  \textbf{X}^*+b)$ \label{linein}
    }
   Compute and save the aiming strategy quality score $QS_i$ for the $\textbf{X}_i$ values using Algorithm \ref{algorithm_aiming_quality_score} \label{linei11}

    Update $D \leftarrow D \cup \{(\textbf{X}_i, QS_i)\}$ \label{linei111}

}

\caption{Data generation procedure}
\label{algorithm_pata_gen}
\end{algorithm}

The inputs include the iteration index $t$, the desired dataset size $N$, parameters $\mu$ and $\sigma$ for generating random values from a Normal distribution, and $a$ and $b$ for a Uniform distribution. It is important to mention that, to guarantee operability and consistency among heliostats within the same row, the values of the aiming $k$ factors are fixed across all heliostats in a given row for each sector of the heliostat field. This simplification aligns with operational constraints and improves the uniformity of the aiming strategy.

In the first iteration, random values of the $k$ factors, i.e., $\textbf{X}_i$, are generated for all heliostats using the specified distributions (line \ref{linei1}). For the second and subsequent iterations, the random values are generated around the optimal solution obtained after solving the optimization problem formulated in Section \ref{sec_opt_problem} (line \ref{linein}). These values are evaluated to compute the quality score $QS_i$ using Algorithm \ref{algorithm_aiming_quality_score} (line \ref{linei11}), and the resulting pairs $(\textbf{X}_i, QS_i)$ are stored in $D$ (line \ref{linei111}).

Finally, it is important to note that the data generation process is inherently parallelizable, enabling the efficient production of a substantial number of samples, often within a matter of seconds.

\subsection{Optimization Problem Formulation}\label{sec_opt_problem}

We propose a surrogate model to optimize the aiming strategy of a CSPT plant by approximating the originally intractable non-linear model. Specifically, we adapt the physical constraints described in Section~\ref{sec:classical_aiming} using an $L$-hidden-layer ReLU NN and its corresponding mixed-integer reformulation. The optimization problem aims to maximize the quality score $QS$, as defined in Section~\ref{sec:aiming_q_score}. The decision variables $\textbf{X} \in \mathbb{R}^{n_{0}}$ represent the aiming factors $k$ of the heliostats, i.e., $\textbf{X} = {x_1, \ldots, x_{n_0}}$, which must remain within the physical limits $x_j \in [k^{\text{min}}, k^{\text{max}}]$ for all $j \in {1, \ldots, n_0}$. The context $\boldsymbol{\Omega}$ includes operational parameters and environmental conditions relevant to the CSPT plant.

To embed the ReLU NN within the optimization framework, we define its components as follows. The NN input variables are $a_j^0$, representing the $j$-th input for all $j \in {1, \ldots, n_0}$. For the subsequent layers, the ReLU activation output for the $j$-th neuron is $a_j^l$, valid for all $j \in {1, \ldots, n_l}$ and $l \in {1, \ldots, L}$. Each neuron in layer $l$ is associated with a continuous variable $z_j^l$, the preactivation value, computed as a linear combination of the previous layer’s activations. This holds for all $j \in {1, \ldots, n_l}$ and $l \in {1, \ldots, L+1}$. The parameters $\textbf{W}_j^l$ and $b_j^l$ denote the NN’s weights and biases, while $M_j^{-,l}$ and $M_j^{+,l}$ are big-$M$ constants ensuring a correct mixed-integer representation. Each neuron also has a binary variable $\sigma_j^l \in \{0,1\}$ indicating whether that neuron is active. The NN’s output corresponds to the computed $QS$, thus linking the decision variables to the performance metric.

Because predictive models like NNs lose accuracy for points far from the training data, especially near the boundary of the feasible region, we use the CH of the dataset as a TR. This restricts the solution search to a reliable interpolation region and avoids unreliable extrapolation. However, a strict CH-based TR can be overly conservative. To mitigate this, we adopt the $\varepsilon$-CH approach from \cite{maragno2023mixed}, which relaxes the TR by allowing solutions outside the CH if they lie within a hyperball of radius $\varepsilon$ around at least one training data point. This relaxation balances predictive accuracy and exploration in high-dimensional spaces, enabling the discovery of better solutions not seen during training.

To incorporate the $\varepsilon$-CH, consider $\hat{\textbf{x}}_i$ as the $i$-th training instance for $i \in {1, \ldots, N}$, and let $\beta \in \mathbb{R}^N$ be a variable that characterizes the CH as the smallest convex polytope containing all $\hat{\textbf{x}}_i$. The relaxation uses $\textbf{s} \in \mathbb{R}^N$ to model deviations from the CH, and solutions remain feasible if they lie within a hyperball of radius $\varepsilon$ centered on at least one $\hat{\textbf{x}}_i$.

Combining the surrogate model, the aiming constraints from Section~\ref{sec:classical_aiming}, and the objective of maximizing $QS$ from Section~\ref{sec:aiming_q_score}, we arrive at problem \eqref{problem_nn_final}:
\begin{subequations}\label{problem_nn_final}
\begin{align}\label{problem_nn_final_eq1}
&\max \quad QS \\\label{problem_nn_final_eq2}
&\text{s.t.} \nonumber\\
&  a_j^0 = x_j,  \quad \forall j \in \{1, \ldots, n_0\}  \\\label{problem_nn_final_eq3}
& QS = z^{L+1}_1, \\\label{problem_nn_final_eq4}
& z_j^l = (\textbf{W}_j^l)^\top \textbf{a}^{l-1} + b_j^l,  \quad \forall j \in \{1, \ldots, n_l\}, \; l \in \{1, \ldots, L+1\}  \\\label{problem_nn_final_eq5}
&  a_j^l \geq z_j^l,  \quad \forall j \in \{1, \ldots, n_l\}, \; l \in \{1, \ldots, L\}  \\\label{problem_nn_final_eq6}
& a_j^l \leq z_j^l - M_j^{-,l}(1 - \sigma_j^l),  \quad \forall j \in \{1, \ldots, n_l\}, \; l \in \{1, \ldots, L\}  \\\label{problem_nn_final_eq7}
& 0 \leq a_j^l \leq M_j^{+,l} \sigma_j^l,  \quad \forall j \in \{1, \ldots, n_l\}, \; l \in \{1, \ldots, L\}  \\\label{problem_nn_final_eq8}
& \sigma_j^l \in \{0, 1\},  \quad \forall j \in \{1, \ldots, n_l\}, \; l \in \{1, \ldots, L\}  \\ \label{problem_nn_final_eq9}
&  k^{\text{min}} \leq x_j \leq k^{\text{max}},  \quad \forall j \in \{1, \ldots, n_0\}  \\\label{problem_nn_final_eq10}
& \sum_{i=1}^{N} \beta_i \hat{\textbf{x}}_i = \textbf{X} + \textbf{s}, \\\label{problem_nn_final_eq11}
& \sum_{i=1}^{N} \beta_i = 1, \\\label{problem_nn_final_eq12}
& \beta_i \geq 0 ,  \quad \forall i \in \{1, \ldots, N\}   \\\label{problem_nn_final_eq13}
& \|\textbf{s}\|_{\infty} \leq \varepsilon. 
\end{align}
\end{subequations}

In \eqref{problem_nn_final_eq1}, we maximize $QS$, accounting for energy uniformity and overall absorption. Constraints \eqref{problem_nn_final_eq2}–\eqref{problem_nn_final_eq8} embed the NN’s architecture into the optimization problem, including the ReLU activations and their mixed-integer formulation. Constraint \eqref{problem_nn_final_eq9} ensures the heliostat aiming factors respect physical limits. Finally, constraints \eqref{problem_nn_final_eq10}–\eqref{problem_nn_final_eq13} define the TR and its relaxation through the $\varepsilon$-CH approach. By ensuring that solutions remain within or near the CH, while allowing for controlled deviations via a hyperball of radius $\varepsilon$, this formulation balances model accuracy, tractability, and solution quality.

\subsection{Iterative Procedure}\label{sec_itera_procedure}

The iterative procedure incrementally refines the heliostat aiming strategy by alternating between data generation, NN training, and optimization. Inspired by the alternating refinement strategy in \cite{kronqvist2023alternating}, this process ensures progressive improvement in both energy collection and flux uniformity. The TR parameter $\varepsilon$ facilitates controlled exploration beyond the current CH of the dataset, thus enabling the discovery of better solutions as iterations proceed.

\begin{algorithm}[tb]
\caption{Iterative optimization with NN refinement}
\label{algorithm_iter_procedure}
\KwInput{
$D$: Initial dataset for NN training\\
$k^{\text{min}}, k^{\text{max}}$: Minimum and maximum $k$ factors bounds for heliostats\\
$L$: Number of hidden layers\\
$M_j^{-,l}, M_j^{+,l}$: Big-$M$ constants for ReLU constraints\\
$N$: Desired dataset size\\
$n_0$: Input layer size\\
$n_l$: Hidden layer size\\
$T$: Number of iterations\\
\(\varepsilon\): TR parameter
}

\KwOutput{Optimal solution $\textbf{X}^*$ for the aiming strategy}

Define an NN with $n_0$ input neurons, $L$ hidden layers, $n_l$ neurons per layer, and ReLU activation functions \label{definition}

\For{$t=1$ \KwTo $T$}{
    Generate a dataset $D$ with $N$ samples using Algorithm~\ref{algorithm_pata_gen} \label{datgen}
    
    Train the NN with the dataset $D$ \label{train_retrain}
    
    Extract \(\textbf{W}_j^l\) and \(b_j^l\) from the NN, and embed them along with $k^{\text{min}}, k^{\text{max}}, L, M_j^{-,l}, M_j^{+,l}, N, n_0$, and $n_l$ into Problem~\eqref{problem_nn_final} \label{embbeding}
    
    Solve Problem~\eqref{problem_nn_final} for different values of \(\varepsilon\) and save the best solution \(\textbf{X}^*\)\label{optimizationsol}

    \If{Stopping criterion is satisfied \label{stopping_crit1}}{
        Break \label{stopping_crit2}
    }
}
\end{algorithm}

Algorithm \ref{algorithm_iter_procedure} summarizes this iterative methodology, beginning with the definition of the NN architecture (line~\ref{definition}). The NN consists of an input layer with $n_0$ neurons corresponding to the heliostat aiming factors, $L$ hidden layers each with $n_l$ neurons, and ReLU activation functions.

In the first iteration, Algorithm~\ref{algorithm_pata_gen} creates an initial dataset $D$ of randomly sampled aiming factors within $[k^{\text{min}}, k^{\text{max}}]$. Each sample’s corresponding $QS$ is computed via Algorithm~\ref{algorithm_aiming_quality_score}. The NN is then trained to approximate the $QS$ based on $D$ (line~\ref{train_retrain}). After training, the NN weights $\textbf{W}_j^l$ and biases $b_j^l$ are integrated into the optimization framework (line~\ref{embbeding}), along with the physical constraints, the ReLU reformulations, and the TR parameter $\varepsilon$. The optimization problem is solved across varying $\varepsilon$ values, and the best $\textbf{X}^*$ is selected as the initial high-quality solution (line~\ref{optimizationsol}).

Subsequent iterations follow the same pattern, but the data generation step focuses on sampling near the current best solution $\textbf{X}^*$. This localized sampling enhances the NN’s fidelity in the region of interest (line~\ref{datgen}), thus improving the accuracy of the surrogate model. Retraining the NN on the updated dataset (line~\ref{train_retrain}) and resolving the optimization problem (lines~\ref{embbeding} and~\ref{optimizationsol}) progressively refine the solution. The parameter $\varepsilon$ allows controlled exploration beyond previously considered regions, encouraging the identification of improved solutions until a stopping criterion (e.g., convergence or maximum iterations) is met (lines~\ref{stopping_crit1} and~\ref{stopping_crit2}).

\begin{figure}[!ht]
    \centering
    \includegraphics[width=\linewidth]{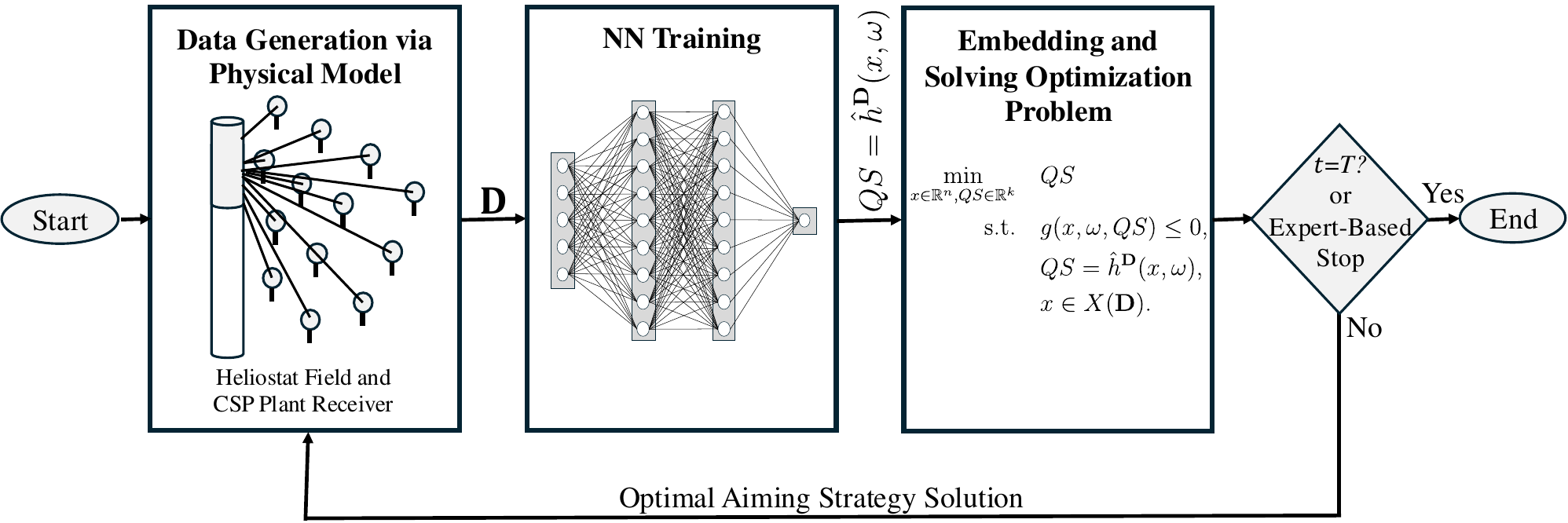}
    \caption{Iterative approach for improving the optimal solution using refined NN training data.}
    \label{fig:iterative_procedure}
\end{figure}

Figure \ref{fig:iterative_procedure} illustrates the iterative process, showing how each cycle refines the dataset, improves the NN approximation, and enhances the quality of the heliostat aiming strategy.

\section{Results}
\label{sec:cases_estudies}

This section presents the results obtained by applying our NN training and optimization approach to the optimal aiming strategy problem at the Dunhuang CSPT plant. In all cases, a single-hidden-layer NN with $50$ neurons was employed. The NN hyperparameters were fixed: Adam was selected as the optimizer, the learning rate was set to $0.0005$, and the batch size was set to $512$. An average of $5,000$ samples were used per iteration, with this number increasing from the first to the last iteration. From a computational cost perspective, an average of $15$ minutes per iteration was required.

For all experiments, a Mac Mini M4 pro with with 24GB RAM was employed. Results were obtained using Pytorch 2.1 \cite{paszke2019pytorch} for NN training and Gurobi 11.0 \cite{gurobi} as the optimization solver for the surrogate problem.

\subsection{Effect of the Distribution Difference Penalty}

Firstly, we analyze the effect of the Distribution Difference ($dd$) penalty $\lambda$ in the results obtained. For this experiment, we run a six-iteration procedure and keep the best result according to the defined $QS$. We consider the equatorial aiming strategy as a base benchmark, where the maximum value of collected energy is obtained, but with a high risk concentration in the receiver.

\begin{figure}[!ht]
    \centering
    \includegraphics[width=0.8\linewidth]{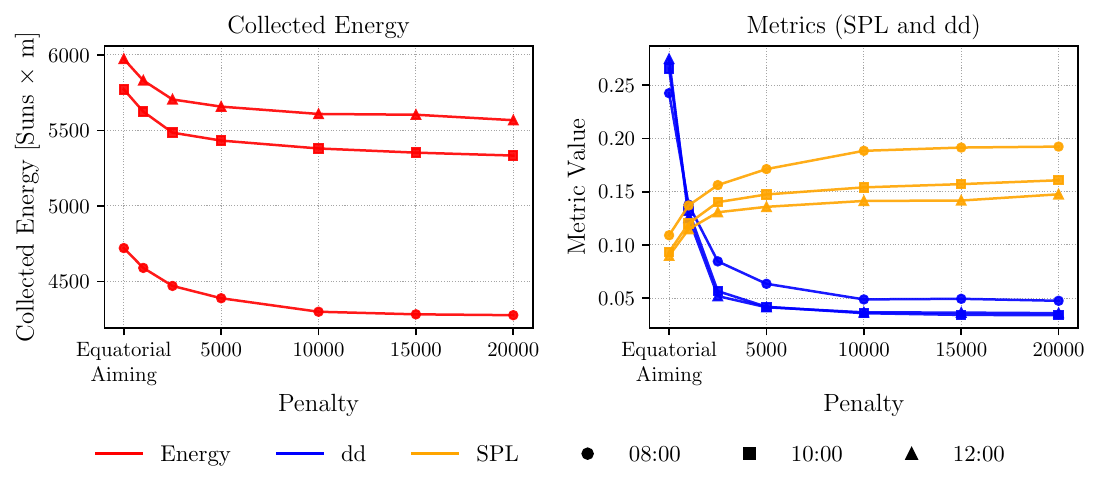}
    \caption{Collected energy and performance metrics versus distribution different penalty ($\lambda$) at 8:00, 10:00, and solar noon.}
    \label{fig:penalty_eval}
\end{figure}

Figure \ref{fig:penalty_eval} presents the interaction between collected energy and key performance metrics, such as $dd$ and $SPL$, as the $\lambda$ penalty varies at three different times of day (8:00, 10:00, and solar noon). On the left, collected energy (in $Suns \times m$) is shown versus the penalty. Under zero penalty, the heliostat field operates under an equatorial aiming strategy that yields relatively high levels of collected energy. As the penalty increases, the overall energy consistently decreases (but not in a high value), indicating that applying constraints to improve the heliostat distribution pattern leads to slightly reduced energy collection values. Although this decline is evident at all three times of the day, the collected energy levels seem to stabilize from a $\lambda$ value of $10,000$.

On the right-hand side, the figure shows the evolution of two performance metrics, $dd$ (in blue) and $SPL$ (in yellow), as the penalty increases. The $dd$ metric, representing the unevenness in the aiming distribution, shows a clear improvement: it starts relatively high with zero penalty and decreases steadily as the penalty grows. Lower $dd$ values indicate a safer, more uniform flux distribution on the receiver surface. In contrast, the $SPL$ metric, which quantifies the fraction of energy that misses the receiver, does not exhibit a similar downward trend. Instead, $SPL$ remains relatively stable as the penalty increases from a value of $5,000$, indicating that while the distribution penalty can significantly improve the flux uniformity (lower $dd$), it does not necessarily translate to a substantial increment in not intercepted energy outside the receiver’s target area.

\subsection{Iterative Procedure Analysis}

We evaluate now the effect of the iterative procedure in the resulting aiming strategy. For this analysis, we fix the penalty value $\lambda$ to $5,000$. This value results in a good trade-off between the collected energy and the distribution difference, as shown in the previous section.

\begin{figure}[!ht]
    \centering
    \includegraphics[width=0.8\linewidth]{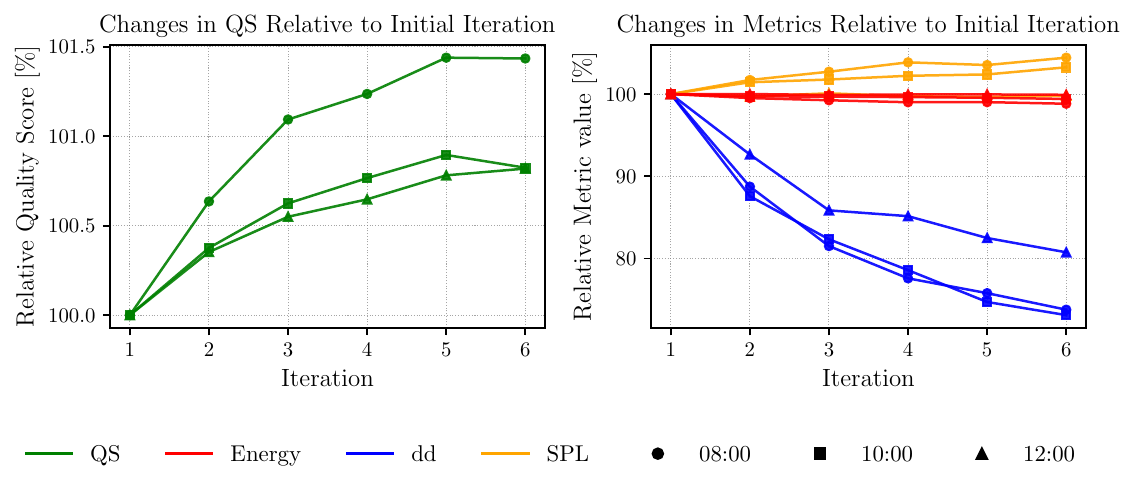}
    \caption{Relative changes in $QS$ and other performance metrics over successive optimization iterations with a fixed $\lambda$ value of $5,000$. Results given at 8:00, 10:00, and solar noon.}
    \label{fig:iteration_eval}
\end{figure}

Figure \ref{fig:iteration_eval} illustrates how the $QS$ and related performance metrics evolve over a series of iterative optimization steps conducted at three different times of day: 8:00, 10:00, and solar noon. Each iteration consists of generating new training data around the previously found optimal solution, retraining the NN surrogate model, and then solving the surrogate optimization to refine the heliostat aiming strategy.

The left-hand side of the figure shows the relative changes in $QS$, represented by a green line. The $QS$ is expressed as a percentage relative to the initial iteration (Iteration 1), serving as a baseline. Across iterations, the $QS$ steadily increases at all times of the day, indicating that the optimization process is successfully improving the overall quality of the aiming strategy, demonstrating the value of iteratively refining the surrogate model and focusing data generation around the previous optimum.

The right-hand side presents the relative changes in three other key metrics, such as the collected energy (red line), $dd$ (blue line), and $SPL$ (yellow line), with each of them also shown as a percentage relative to the initial iteration. While the collected energy and spillage values remain relatively stable across the iterations, $dd$ exhibits a substantial decrease. A lower $dd$ corresponds to more uniform flux distributions on the receiver, suggesting that the iterative process not only improves the overall quality of the solution (higher $QS$) but also leads to more balanced and potentially safer flux patterns. Importantly, these trends are consistently observed at 8:00, 10:00, and solar noon, highlighting that the approach is robust to varying solar conditions.

\subsection{General results and Comparisons}

This section presents and compares the flux distributions obtained using the optimized aiming strategy (referred to as NN+Opt) against a reference case, the Sweep approach proposed in \cite{Sanchez-Gonzalez2018}. Results are shown at three representative times during the equinox day: 08:00, 10:00, and 12:00 (noon) solar hours.

\begin{figure}[!ht]
    \centering
    \begin{subfigure}
        \centering
        \includegraphics[width=\linewidth]{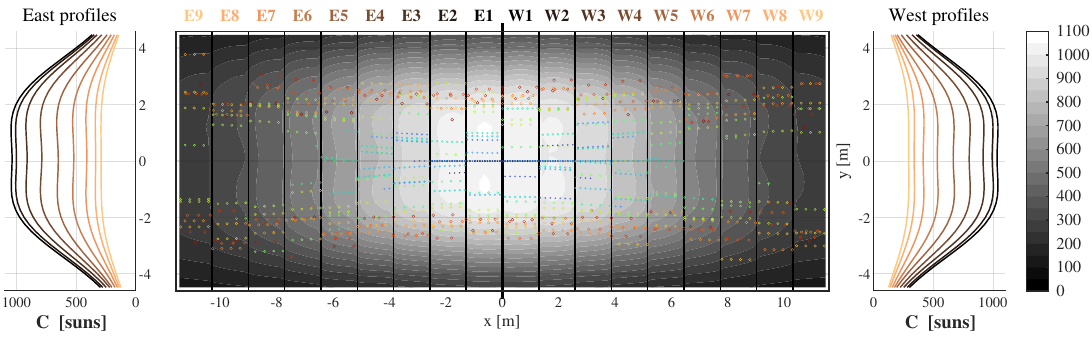}
        \label{fig:cmann_eq12}
    \end{subfigure}

    \begin{subfigure}
        \centering
        \includegraphics[width=\linewidth]{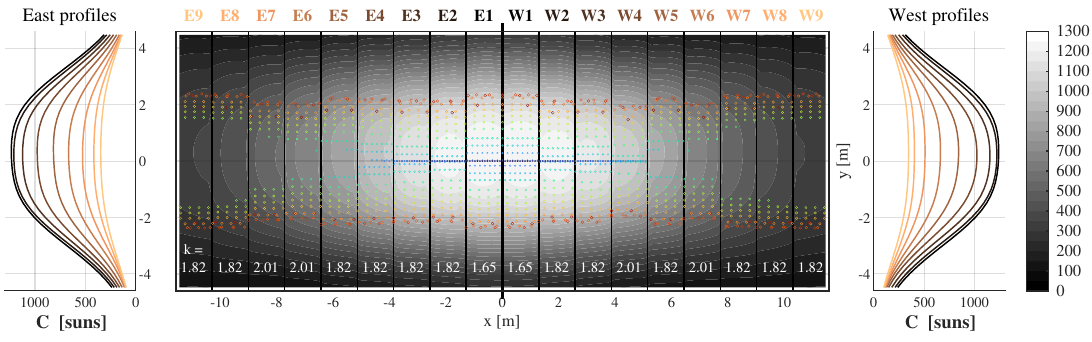}
        \label{fig:cmflat_eq12}
    \end{subfigure}
    
    \caption{NN+Opt (up) and Sweep (down) at solar noon. Maps of concentration and aim points, along with concentration profiles per panel.}
    \label{figCM12h}
\end{figure}

Figure~\ref{figCM12h} illustrates the flux maps at equinox solar noon, comparing NN+Opt (top) and Sweep (bottom). Each heliostat’s aim point is marked with a colored circle as defined in Figure~\ref{figHF}, and the background grayscale contours indicate the concentration ratio $C$ on the receiver. To the left and right of these maps, the mean vertical concentration profiles $C^V_{p,v}$ for the eastern and western panels, respectively, are plotted. The profile color coding corresponds to the panels highlighted above the maps.

Both flux maps at solar noon show symmetry with respect to the receiver’s equator ($y=0$) and the north-south axis ($x=0$), as expected due to the nearly symmetric heliostat field layout at solar noon. While both methods achieve symmetry, the flux distribution obtained from the NN+Opt strategy exhibits a substantially flatter profile across each panel compared to that produced by the Sweep approach. This improvement can be clearly seen in the concentration profiles: the NN+Opt results yield less pronounced peaks and more uniform distributions across the receiver’s vertical dimension. Table~\ref{tabComparison} reports a 52.3\% reduction in the distribution difference $dd$ for the NN+Opt solution at 12:00 compared to Sweep.

\begin{table}[!ht]
\centering
\resizebox{\textwidth}{!}{%
\begin{tabular}{l *{6}{r}}
\toprule
 & \multicolumn{2}{c}{8:00} & \multicolumn{2}{c}{10:00} & \multicolumn{2}{c}{12:00}\\
\cmidrule(lr){2-3} \cmidrule(lr){4-5} \cmidrule(lr){6-7}
 & \multicolumn{1}{c}{NN+Opt} & \multicolumn{1}{c}{Sweep} & \multicolumn{1}{c}{NN+Opt} & \multicolumn{1}{c}{Sweep} & \multicolumn{1}{c}{NN+Opt} & \multicolumn{1}{c}{Sweep} \\
\midrule
Collected Energy    & 4388.356 (1.02\%) & 4344.235 & 5432.281 ($-1.8\%$) & 5533.687 &  5657.112 ($-2.2\%$) & 5785.468 \\[3pt]
Distribution Diff.  &  0.063 ($-33.7\%$) & 0.095 &  0.042 ($-43.2\%$) & 0.074 & 0.041 ($-52.3\%$) & 0.086 \\[3pt]
$SPL$                 & 0.171 ($-12.8\%$) & 0.196 & 0.147 (10.5\%) & 0.133 & 0.136 (15.3\%) & 0.118 \\[3pt]
Max Suns            & 860.6 ($-9.2\%$) & 947.4 & 1042.3 ($-7.5\%$) & 1127.2 & 1053.2 ($-8.9\%$) & 1155.5 \\
\bottomrule
\end{tabular}
}
\caption{Comparison of Sweep and NN+Opt approaches for different metrics and times.}
\label{tabComparison}
\end{table}

The flatter flux profile achieved by NN+Opt reduces peak concentrations from 1156 suns (Sweep) to 1053 suns (NN+Opt). Although this reduction alleviates hot spots and potential thermo-mechanical stresses, it comes at the cost of a slight reduction in collected energy (2.2\% at 12:00). This trade-off also slightly increases the spillage loss ($SPL$) by approximately 1.8 percentage points, from 11.8\% (Sweep) to 13.6\% (NN+Opt).

The Sweep approach’s aiming strategy relies on fixed aiming zones and uniform $k$-factors within discrete regions of the heliostat field. As a result, the aim points cluster around these pre-defined levels, preserving clear symmetry and uniformity in the aiming pattern. In contrast, the NN+Opt method does not rely on discrete aiming levels, allowing for a continuous adjustment of $k$-factors for each heliostat. While this flexibility breaks the strict geometric symmetry seen in the Sweep results, it provides a more finely tuned flux distribution and improved receiver thermal management.

\begin{figure}[!ht]
    \centering
    \begin{subfigure}
        \centering
        \includegraphics[width=\linewidth]{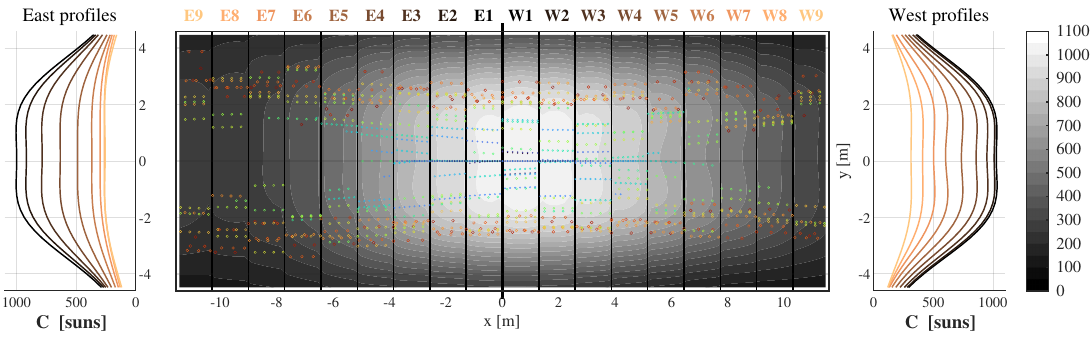}
        \label{fig:cmann_eq10}
    \end{subfigure}

    \begin{subfigure}
        \centering
        \includegraphics[width=\linewidth]{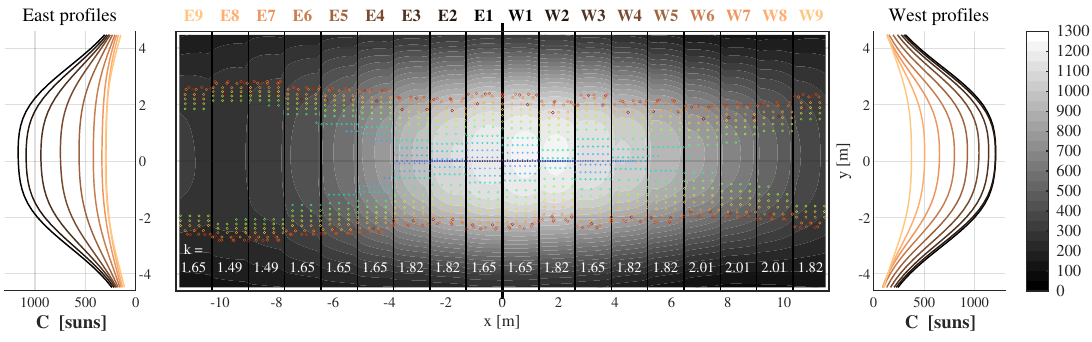}
        \label{fig:cmflat_eq10}
    \end{subfigure}
    
    \caption{NN+Opt (up) and Sweep (down) at 10:00, solar time. Maps of concentration and aim points, along with concentration profiles per panel.}
    \label{figCM10h}
\end{figure}

Figure~\ref{figCM10h} shows the flux maps and profiles at 10:00 solar time. During this period, the receiver’s west side (panel W1) dominates the flux due to the sun’s position. The NN+Opt approach once again achieves a lower peak concentration (1042 suns) compared to the Sweep solution (1127 suns), reducing the risk of hot spot formation. The NN+Opt solution also flattens the flux distribution more effectively, as reflected in the $dd$ values (0.042 for NN+Opt vs. 0.074 for Sweep). Although this improved uniformity slightly reduces collected energy by 1.8\%, the increase in spillage loss is limited to about 1.4 percentage points.

\begin{figure}[!ht]
    \centering
    \begin{subfigure}
        \centering
        \includegraphics[width=\linewidth]{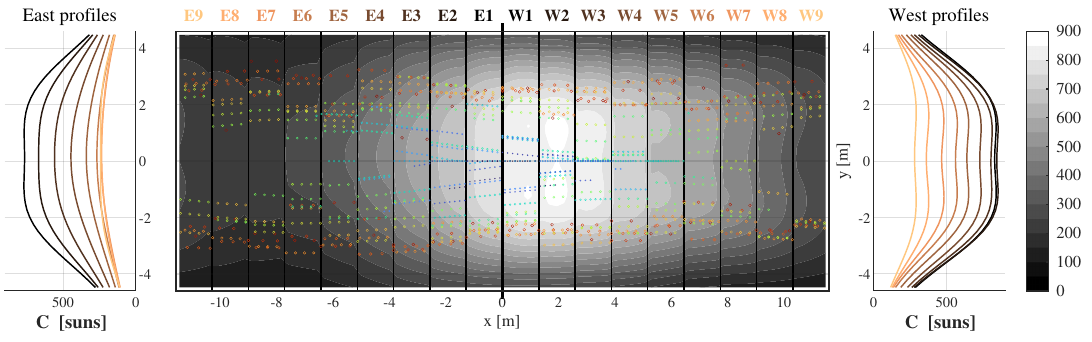}
        \label{fig:cmann_eq8}
    \end{subfigure}

    \begin{subfigure}
        \centering
        \includegraphics[width=\linewidth]{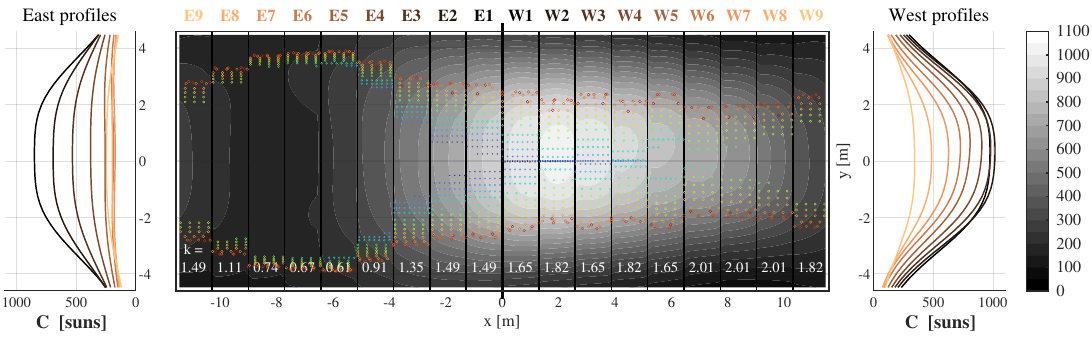}
        \label{fig:cmflat_eq8}
    \end{subfigure}
    
    \caption{NN+Opt (up) and Sweep (down) at 08:00, solar time. Maps of concentration and aim points, along with concentration profiles per panel.}
    \label{figCM8h}
\end{figure}

Figure~\ref{figCM8h} presents the results at 08:00 solar time, a period with a more pronounced east-west asymmetry due to the sun’s low position in the sky. Here, the peak concentration shifts towards panel W2. The NN+Opt method reduces peak flux (596 suns) relative to Sweep (614 suns) and significantly flattens the distribution, with $dd$ dropping from 0.095 (Sweep) to 0.063 (NN+Opt). Notably, at this earlier hour, NN+Opt not only improves flux uniformity but also collects slightly more energy than Sweep (+1.02\%), while simultaneously reducing spillage from 19.6\% to 17.1\%.

Overall, these results demonstrate that the NN+Opt strategy consistently achieves flatter flux distributions, reduces peak flux levels, and generally lowers the distribution difference metric. While this often leads to marginally reduced energy collection and slightly increased spillage losses at mid-morning and noon, the approach provides a clear benefit in terms of mitigating hot spots and potential receiver damage. At early morning times, NN+Opt can even improve energy collection efficiency. This balance between thermal load distribution and energy yield is crucial for optimizing receiver performance and longevity.

\section{Conclusions}
\label{sec:conclu}

This work presents a novel data-driven framework that integrates constraint learning, neural network surrogates, and mathematical programming to optimize heliostat field aiming in CSPT plants. By capturing nonlinear heliostat-to-receiver flux interactions from simulation data, the proposed approach enables continuous, flexible adjustments of aiming factors, resulting in significantly more uniform flux distributions. Using a flat distribution as the baseline, the distribution difference is reduced by up to $50\%$ compared to the classic sweep method. Furthermore, peak solar concentrations are lowered by nearly $10\%$, therefore mitigating the risk of hotspots and prolonging receiver life, but without substantially compromising overall energy collection.

A key advantage of the method lies in its iterative refinement, in which newly sampled data around the current solution iteratively improve the predictive accuracy of the neural network surrogate. Trust regions maintain the solution search within well-represented areas, preventing model extrapolation into unreliable domains. This careful orchestration of learning and optimization leads to systematic enhancements in flux uniformity and receiver performance, as demonstrated in the case study considered.

Building on these results, future work could focus on planning non-static aiming strategies over finer time scales of minutes, considering continuously updated operational parameters and thermal constraints that mirror real-world variability. Such dynamic real-time adjustments would enhance the system's response to rapidly changing solar conditions and temperature profiles, ultimately improving operational efficiency and the longevity of components. In addition, incorporating robust and stochastic formulations into the optimization models would help manage uncertainties in solar resource forecasts and component reliability, ensuring that the proposed methodology remains effective and resilient under diverse operating scenarios.

\section*{Credit authorship contribution statement}


\textbf{Antonio Alc\'antara:} Conceptualization, Methodology, Software, Data Curation, Visualization, Writing - Review \& Editing. \textbf{Pablo Diaz-Cachinero:} Conceptualization, Funding acquisition, Methodology, Software, Writing - Original Draft, Visualization, Supervision. \textbf{Alberto S\'anchez-Gonz\'alez:} Conceptualization, Methodology, Software, Writing - Original Draft, Visualization, Funding acquisition, Supervision. \textbf{Carlos Ruiz:} Conceptualization, Funding acquisition, Writing - Review \& Editing.

\section*{Declaration of competing interest}

The authors declare that they have no known competing financial interests or personal relationships that could have appeared to influence the work reported in this paper.

\section*{Acknowledgements}
This work has been supported by the Madrid Government (Comunidad de Madrid - Spain) under the Mutiannual Agreement with UC3M (SOLAROPIA-CM-UC3M). The authors gratefully acknowledge the financial support from MCIN/AEI/ 10.13039/ 501100011033, project PID2023-151013NB-I00, and the FPU grant (FPU20/00916). 

\section*{Declaration of generative AI and AI-assisted technologies in the writing process}
During the preparation of this work the authors used ChatGPT in order to improve language and readability. After using this tool/service, the authors reviewed and edited the content as needed and take full responsibility for the content of the publication.


\bibliographystyle{elsarticle-num}
\bibliography{biblio}


\end{document}